# Measurement of filling factor 5/2 quasiparticle interference: observation of e/4 and e/2 period oscillations

R.L. Willett*, L.N. Pfeiffer, K.W. West
Bell Laboratories, Alcatel-Lucent
600 Mountain Avenue
Murray Hill, New Jersey, 07974 USA
Correspondence to: rlw@alcatel-lucent.com

ABSTRACT:

A standing problem in low dimensional electron systems is the nature of the 5/2 fractional quantum Hall state: its elementary excitations are a focus for both elucidating the state's properties and as candidates in methods to perform topological quantum computation. Interferometric devices may be employed to manipulate and measure quantum Hall edge excitations. Here we use a small area edge state interferometer designed to observe quasiparticle interference effects. Oscillations consistent in detail with the Aharonov-Bohm effect are observed for integer and fractional quantum Hall states (filling factors $\nu$ =2, 5/3 and 7/3) with periods corresponding to their respective charges and magnetic field positions. With these as charge calibrations, at $\nu$=5/2 and at lowest temperatures periodic transmission through the device consistent with quasiparticle charge e/4 is observed. The principal finding of this work is that in addition to these e/4 oscillations, periodic structures corresponding to e/2 are also observed at 5/2 $\nu$ and at lowest temperatures. Properties of the e/4 and e/2 oscillations are examined with the device sensitivity sufficient to observe temperature evolution of the 5/2 quasiparticle interference. In the model of quasiparticle interference, this presence of an effective e/2 period may empirically reflect an e/2 quasiparticle charge, or may reflect multiple passes of the e/4 quasiparticle around the interferometer. These results are discussed within a picture of e/4 quasiparticle excitations potentially possessing non-Abelian statistics. These studies demonstrate the capacity to perform interferometry on 5/2 excitations and reveal properties important for understanding this state and its excitations.

\body

Experimentally the fractional quantum Hall (FQH) state at 5/2=$\nu$ filling factor is anomalous in that it occurs at an even denominator quantum number[1]. The state is fragile: it displays a weak quantum Hall effect, requiring temperatures for observation substantially lower than the principal odd denominator states at $\nu$ = 1/3 and 2/3. It has been proposed as either a spin polarized (Moore-Read Pfaffian [2]) or non-spin polarized (Haldane-Rezayi[3]) paired composite fermion state. The fundamental quasiparticle excitations are expected to be charged e/4, and for the Pfaffian state these quasiparticles are to obey non-Abelian statistics. These non-Abelian states may display utility in topological quantum computational schemes [4]

To the end of determining the charge, and more importantly the statistics of the quasiparticles, interference devices and their function in displaying the Aharanov-Bohm effect at 5/2 have been described theoretically[5-9]. Interference devices are typically constructed with nominally two adjacent quantum point contacts (qpcs), each able to variably transmit current, and a confinement area between the qpcs (see schematic in Figure 1). Through splitting of the current at the qpcs, redirection of quantum Hall edge currents around the confined area occurs. Interference of that encircling edge with the other backscattered current is expected to yield oscillatory resistance across the device for changing confinement area, with period dependent upon the edge current charge. The encircling quasiparticle statistics may also be assessed by examining the interference pattern if the confinement area holds determinable localized quasiparticles [5-9]. This interferometric method is attractive in that for a given device it can be applied to a series of integer and fractional quantum Hall states for verification of its operation through measurement of their respective charges, including the charge at 5/2 filling factor.

Edge transport interference in integer quantum Hall (IQH) systems has been reported for both accidental and intentional area confinements [10-11]. In Mach-Zehnder interferometer designs [12-13] oscillatory transmission is observed for both B-field sweep and enclosed area sweep at IQH states. Fractional quantum Hall effect interference measurements have been more elusive. An extensive set of studies by Goldman et al [14] have reported well defined periodic structures in interferometer transmission near FQH filling factors. Later similar studies by Godfrey, et al [15] over a filling factor range of both IQH and FQH states have shown periodic conduction with period linearly increasing with magnetic field. These results [14-15] are qualitatively similar, with both interference devices fabricated as etched and gated structures on the few micron size scale. Such device processing supports the more robust IQH systems, but may adversely affect the smaller gapped FQH states [16]. In theoretical analysis of these results and interference devices in general, Rosenow and Halperin [17] point out that periodic transmission results may be attributable to Coulomb blockade effects rather than edge state interference. This is of particular significance if the device fabrication adversely affects electron correlations, emphasizing the goal of developing an interferometric device that can accommodate the less robust FQH states.

Here we report interferometry studies using devices particularly designed for accommodating fragile FQH states, specifically the 5/2 state. FQH state confinement in an interferometer geometry is achieved here using only top gate structures on ultra-high mobility, high density samples, where the gate structures have minimal negative impact on the 2D electron quality. These top gate structures are designed to allow independent control of the size of the encircled area and the transmission qpcs, thus allowing area modulation to be used to assess quasiparticle interference. With these devices, periodic conduction oscillations are observed for both magnetic field sweeps and lateral confining gate sweeps. The oscillatory transmission observed in these studies occurs over small path lengths (~few microns) and demonstrates phase noise. The focus of this study is the finding of periodic conduction features at $\nu=5/2$ for lateral confining gate sweeps. At lowest temperatures oscillation periods corresponding to charge e/4 are observed, using as calibration periodic conduction at $\nu=2$ assigned to charge e and with resistive oscillations at $\nu=5/3$ and 7/3 having period consistent with charge e/3. The principal finding of this study is that at 5/2 filling in addition to the e/4 oscillations, at lowest temperatures oscillations of half the e/4 period corresponding to e/2 can also be observed. Properties of these oscillations at 5/2 are shown here, including amplitude, relative prevalence and preliminary temperature dependence. Among possible causes, this effective e/2 period may empirically reflect the presence of an e/2 quasiparticle, or may be due to a double pass of the e/4 quasiparticle around the interferometer. These results are considered within the important possible picture of *non-Abelian* e/4 quasiparticle statistics.

Operation of an interferometric device in the quantum Hall regime can follow in two modes, examining transmission while either sweeping magnetic field or sweeping device area. Phase accumulation can occur by encircling localized charge or by encircling magnetic flux quanta. In quantum Hall systems application or retraction of magnetic field changes both the magnetic flux density and the quasiparticle number [18], therefore influencing the phase accumulation in two ways. This suggests that B-field sweeps produce a somewhat complicated interferometric scheme for analysis. However, staying at fixed magnetic field but changing the encircled area size via a side gate provides a simpler method for analysis. By changing the encircled area size via a side gate the number of encircled magnetic flux quanta is changed inducing Aharonov-Bohm oscillations. The period of these A-B oscillations indicates the current carrying charge. Another contribution to phase accumulation will occur in this method if the change in area via the change in side gate voltage also includes a localized quasiparticle in the changed area. However, this density of localized quasiparticles is typically substantially below that of the magnetic flux density [18], with the sweep over magnetic flux lines consequently the dominant source for resistive oscillations. The side gate sweep is therefore the preferred technique given its simpler interpretation. While a complete theoretical understanding of edge state interference is presently emerging (B.I. Halperin, A. Stern, private communication), it has been demonstrated experimentally[12-13] that the area change achieved using side gate bias produces Aharonov-Bohm oscillations in which the period directly reflects the charge of the edge channels. Attempts to assess the 5/2 quasiparticle charge and statistics through interferometry are therefore best served by employing a device capable of adjusting the area with a side gate while controlling the transmission independently using separately controlled quantum point contacts.

The operation of the interferometer used here is based upon generating edge current paths that are schematically shown in Figure 1a: current from left contact 1 is split at qpc a, with some part of that current traversing the encircled central area A, and upon reflection at qpc c returns to qpc a via the right edge where it interferes with the current backscattered at a. The central lithographic area $A_L$ can be adjusted resulting in encircled area A using the channel gates b by adjusting the channel voltage $V_s$. The resistance across the device ($V_{3-4}$) should demonstrate oscillations upon interference with period $\Delta B=\phi/A=(h/e^*)/A$ for B-field sweeps or period $\Delta A=\phi/B=(h/e^*)/B$, $\Delta A\sim V_s$ for side gate (gate b) sweeps, again neglecting quasiparticle number change for each as described above. The device details are presented in the supporting information, section I; methods and device operation.

Bulk transport and transport through the interferometer are shown in Figures 1c and 2a respectively. Details of measurements are provided in supporting information section I. The bulk transport shows the lowest Landau level series of fractional states at the high B-field side of $\nu=2$, and a well developed 5/2 state. The longitudinal resistance through the device, $R_L$, is determined using current contacts 1 and 2, and voltage drop is measured along 3 to 4 [19]. With complete depletion under the gates at ~-2.5V, the qpcs and the central channel are biased negatively beyond this value for transmission that collectively promotes observation of interference effects and to adjust total area A, respectively. The $R_L$ data show persistence of the larger gap FQH state zero resistances through the device, and transport at 5/2 shows roughly 200Ω of backscattering or reflection at that filling factor at base temperature.

Oscillations in the longitudinal component $R_L$ are observed in our measurements for both B-field sweeps and side gate Vs sweeps. First B-field sweep results are reviewed. For B-field sweeps periodic structures can be observed over a range of FQH and IQH states. Periodic structures can be coarsely observed over most of the B-field range shown in Figure 2a inset. Close examination of the resistive oscillations near ν=2 is provided by Figure 2a. Here the oscillations are of period ΔB=230G +/- 20G, which corresponds to an effective encircled area A of ~ 0.2 $\mu m^2$, much smaller than the lithographically defined area. This finding is consistent with a small density gradient from depletion to full density as might be expected with our top gate configuration. This gradual change from depletion to full density would imply that the inner edge state is well removed from the lithographic edge and therefore encloses a substantially smaller area; area A << area $A_L$. Using the channel gate b, the encircled area A can be adjusted. For a change in this side gate Vs of ~-3V, a 20% smaller area a can be generated [16], which resulted in an oscillation period that commensurately increased by 20%. The observation of oscillations with similar B-field period over a range of filling factors is consistent with past experimental results [14]. Given this period change with adjustment of area A, it is estimated that a side gate voltage adjustment of approximately 1mV will change the area corresponding to one flux quantum at B-fields near ν=2.

We turn now to observation of resistance oscillations for variations of the side gate voltage (Vs) for different filling factors. As noted above a side gate bias changes the area and so changes the number of enclosed flux quanta such that the oscillation period obeys ΔVs~ ΔA~(h/e*)/B. To establish the relationship of period to side gate bias, the interferometer was operated near filling factor 2, 5/3 and 7/3. Along the high field side of ν=2 near $R_L$~200Ω, side gate bias is swept and $R_L$ is measured; see typical traces in Figures 2b and 2d. All swept gate data displayed in this study correspond to measurements taken on the high B-field side of the indicated filling factor. Sequences of periodic oscillations are observed, and again come in runs or sets, rarely with more than 10 oscillations, but with multiple sets discernible as the side gate is swept. Such sets are marked in Figures 2b and d with vertical lines. The vertical lines within each set of oscillations have a single valued separation that is the measured period for that filling factor as determined by fast Fourier transform (FFT) of the side gate sweep $R_L$ measurement. FFT spectra of the side gate sweep data at ν=2 are shown in Figure 2c. These multiple sets of oscillations are the common observation in these measurements of $R_L$ vs. $V_S$ for the interferometer. The fact that long, continuous strings of oscillations are not observed may indicate that a phase disruptive noise plays a crucial role in the propagation of edge states. In spite of the phase disruptions, the Fourier analysis exposes a peaked frequency spectrum, as shown in panel 2c. The peak values of the spectra are at the frequencies marked by the vertical lines in the $R_L$ data of Figures 2b and 2d. All FFT spectra presented in this study are extracted from the entire side gate voltage range displayed in their respective $R_L$ vs. $V_S$ data figures.

The magnetic field was then adjusted so that the bulk and device filling factor are near 5/3 or 7/3, where Vs is then swept. The $R_L$ vs. Vs spectra at these respective filling factors are displayed in the lower traces of Figure 2b and d. Here the predominant period observed at 5/3 and 7/3 for several sets of oscillations in each trace is clearly larger than that of their respective traces near $\nu=2$. This is verified by the Fourier spectra in Figure 2c comparing filling factor 2 with 5/3 or 7/3 for the two sample preparations. Using the period from the above $\nu=2$ traces an oscillation period is derived for quasiparticle charge of $e^*=e/3$ given the period finding near $\nu=2$ and also considering the B-field correction for the change in flux density, $B(\nu=2)/B(\nu=5/3 \text{ or } 7/3)$. This expected period is marked in the 5/3 and in the 7/3 traces of Figures 2b and d by vertical lines of thus prescribed separation: good agreement is observed between the large oscillatory features of the trace and the expected period for charge e/3 Aharonov-Bohm oscillations. Again, this is supported by the Fourier spectra of the various filling factors (Figure 2c), with the peak values in the FFTs in agreement with the expected periods and corresponding to the marked period lines. The oscillations are observed in sets or packets that are out of phase with other packets in the same Vs sweep, as the runs of oscillations appear to display disruptions in phase. Given these disruptions in phase, however, the predominant oscillation features in each trace can be assigned almost completely to the A-B period of e/3. Note that in Figure 2d, excluding a few minor peaks at the noise level of the measurement, all major features within the trace fall under three phase shifted sets of the appropriate 7/3 period. These marked periods at 5/3 and 7/3 are placed assuming the evident phase disruptions, but the periods used are quantitatively consistent with the FFTs of the data shown in panel 2c. In each case the periods for the different filling factors derived from the FFT peak frequencies are in excellent agreement with their respective expected Aharonov-Bohm periods.

The small effective area of the biased interferometer may particularly allow interference of fractional state excitations with small coherence lengths, as could be the case for 5/3 and 7/3. The largest path length allowable in our devices, the lithographically defined perimeter, is roughly 5 μm, while the path length corresponding to the effective area derived from B-field sweeps of 0.2 $\mu m^2$ is about $2\pi r \sim 1.5$ μm.

With this validation of the interferometer operation for determining quasiparticle charge, measurements are then made near $\nu=5/2$, with important results. Figure 3 shows device resistance as Vs is swept at multiple filling factors including $\nu=5/2$ for a sample following two different cool-downs from room temperature to base temperature. From each cool down the high mobility 2D system displays slightly different properties in standard bulk transport measurements, and likewise displays different properties in standard transport through the interference device. For this reason we display two typical $R_L$ vs. Vs trace sets in the Figure to provide a demonstration of both the persistence of the Aharonov-Bohm effects and the variability in the details of the traces for samples prepared differently. Additional trace sets are provided in the supporting information sections II and III, including a discussion of the high reproducibility of oscillatory features in Vs sweep data. Each panel shows oscillations near $\nu=2$ where the predominant period there is then translated to the expected period for oscillations at either $\nu=5/3$ or $\nu=7/3$: these expected periods are marked in each $\nu=5/3$ or 7/3 trace with good correspondence to oscillatory features. These are the controls for 5/2. FFTs of the side gate sweep measurements of $R_L$ near $\nu=2$ and $\nu=5/3$ or $\nu=7/3$ are shown in insets in the bottom panels. The peaks in the frequency spectra demonstrate quantitative agreement of the respective filling factors for expected AB oscillations.

Figure 3 panels show data taken near 5/2, with evidence for e/4 charge present in both sample preparations. In each panel at 5/2 large period oscillations are observed at these low temperatures (~25mK). These prominent oscillations correspond to the e/4 quasiparticle interference. They demonstrate a good agreement in period to the period derived for e/4 quasiparticles from the ν=2 data and the data at 5/3 and 7/3. The ν=2 and 5/3 or 7/3 measured period is used to derive the anticipated period for oscillations at 5/2 for a quasiparticle charge e*=e/4, with respective periods marked by vertical lines. Note that both the appropriate charge and the magnetic field scaling are necessary to achieve proper fit to the oscillations. Also apparent in these traces is smaller amplitude, smaller period oscillations: these will be addressed below. Sets of less than ten oscillations with these large e/4 periods are observed, again consistent with the limited range oscillation sets for ν=2 and ν=5/3 or 7/3. Fast Fourier transform data shown in the bottom panels demonstrate peaks in the spectra at the frequency expected for e/4 given the FFT spectral peaks at filling factors 2, 5/3, and 7/3. The narrow FFT peaks corresponding to the respective thirds filling factor data provide specific determination of the expected 5/2 FFT peak position and are used in their calculation. Further examples of 5/2 data are shown in supporting information section III.

These large period oscillations are consistent with interference of charge e/4 quasiparticles, given the correspondence in detail of charge scaling (e to e/4) and magnetic field scaling (from near ν= 2 to near ν= 5/2) as expected for Aharanov-Bohm oscillations. Note that the observed oscillation periods near filling factors 5/2, 7/3, 2, and 5/3 do not respectively progress monotonically in B-field. Previous interferometer results[13] potentially attributable to Coulomb blockade possessed monotonic B-field dependence of the oscillation period, distinctly different from the results shown here] These prominent oscillatory features with a period corresponding to e/4 provide substantial support for the model that the elementary excitation at 5/2 has charge e/4.

A critical finding in this study is the presence near filling factor 5/2 of prominent oscillations displaying *two* different periods, the period consistent with charge e/4 and a period consistent with e/2. In the traces of Figure 3, in addition to the oscillations of e/4 period other distinct oscillations are apparent with a shorter period that corresponds to e/2, and are marked accordingly with appropriate vertical lines. In the Figure these shorter period oscillations appear here in sets or runs of oscillations adjacent sequentially in side gate voltage to the sets of e/4 oscillations. The presence of the e/2 period is quantified by the presence in the FFT spectra of a peak corresponding to e/2, shown in the Figure 3 bottom panels. Further examples of such e/2 period oscillations are provided in supporting information section III, including isolation of the e/2 oscillations and FFT demonstration of that period. Figure 3 demonstrates that these e/2 period oscillations can be present at the lowest available temperatures of this study in side gate sweeps and at side gate voltages adjacent to those revealing e/4 oscillations. This finding is further demonstrated in Figure 4a and 4b, showing in other sample preparations additional measurements of $R_L$ through the interferometer where the side gate voltage Vs is swept. Again, in these traces two periods are observed, one period in size corresponding to e/4 by reference to the ν=2 period, and another of half that period, which is an effective e/2 period. This data further suggest the possibility that e/4 and e/2 periods may coexist in some fashion as the side gate is swept at the lowest temperatures available in this study. As in Figure 3, the e/2 period oscillations are in general of lower amplitude than the e/4 oscillations.

Beyond these properties of the e/2 and e/4 oscillations, preliminary temperature dependence is displayed in Figure 4. The top two panels show typical prevalence of the e/4 and e/2 periods at the base temperature near 5/2 filling factor. The data include FFT spectra demonstrating the presence of peaks corresponding to both e/4 and e/2. The influence of temperature on the presence of this shorter period e/2 oscillation is tested in panel c, in which the sample temperature is changed during a side gate sweep. The sample is first held at roughly 30mK, where large period oscillations are observed corresponding to e/4 as marked. The sample temperature is then abruptly increased to near 150mK, and the shorter period consistent with e/2 is exposed. These results indicate that at higher temperatures the predominant A-B period may become e/2 with diminishing occurrence of e/4 oscillations. This is further examined by the data in panel c. Here three different side gate sweeps are performed at progressively higher temperatures. These sweeps are measured with a low drive current of 2nA; other data collected at higher drive current is shown in supporting information section IV. At lowest temperature periods consistent with e/4 and e/2 are present. At the higher temperature of 130mK a greater prevalence of e/2 oscillations is observed: at this temperature the 5/2 state in bulk still demonstrates a distinct minimum in Rxx. In these samples the activation energy in bulk measurement is approximately 150mK. At the next higher temperature (610mK) shown in the Figure the overall amplitude of the $R_L$ measurement is substantially diminished, with only a background noise apparent. These 5/2 data suggest that as temperature is increased, the e/2 oscillations may become the prevailing feature before elimination of the quantum Hall state above the temperature corresponding to the activation energy. The data of Figure 4 and the supporting information provide only a preliminary indication of the e/4 and e/2 temperature dependence, with further temperature studies presently underway.

The principal findings of this study using a small area interferometer in which the area is changed with a side gate are i) results consistent with interference of fractional quantum Hall edge currents are observed as supported by data of transmission oscillations with period corresponding to both charge and magnetic field positions at filling factors 2, 5/3, and 7/3, ii) in the model of interference of edge currents, at filling factor 5/2 transmission oscillations of period appropriate to charge e/4 are demonstrated, and iii) near 5/2 an additional oscillation period corresponding to e/2 is prominent in the side gate sweep data.

The finding of e/4 period oscillations at 5/2 is consistent with the present theoretical understanding of the 5/2 state excitations [2, 3] and their common predicted charge of e/4. This result, as with previous measurements deducing 5/2 excitation charge [19-20], provides a consistency with the paired state pictures of the Moore-Read state and the Haldane-Rezayi state, yet cannot discriminate between these models. Additionally, this charge assessment does not determine the underlying statistics of the excitation.

The origin of the e/2 oscillations at lowest temperatures can include empirically two fundamental possibilities: a) the oscillations represent the presence of a quasiparticle of e/2 charge, or b) the oscillations are the result of an e/4 quasiparticle completing two full passes or laps around the interferometer before the interference occurs; the encircled area is doubled (the number of encircled magnetic flux are doubled) and therefore the period is halved. The first possibility, that the smaller e/2 period directly reflects the presence of e/2 quasiparticles, is not supported by early theory which describes the elementary excitations as e/4 [2, 3, and 5, 6]. However, the possibility of e/2 excitations has been broached in more recent work.[22] A viable picture of the presence of an e/2 quasiparticle must also include a rational for a coexistence with e/4 quasiparticles and how either e/2 or e/4 period is expressed in the data sequentially, as observed in the data, but not in general simultaneously.

The second possibility in which e/4 quasiparticles produce both e/2 and e/4 periods, the e/2 period due to e/4 accomplishing two laps around the interferometer, has the fundamental constraint that the coherence length of the e/4 quasiparticle must be sufficient to allow completion of the longer double pass. This may be the case given the small perimeter of the interferometer and the apparent preservation of the 2D electron system high quality in the device. In addition, the generally lower amplitude of the e/2 oscillations supports this possibility. An argument against this multiple pass picture for e/2 is that as the temperature is increased, the e/2 oscillations may become *more* prevalent, not less. Heuristically it might be expected that the larger path length of the double traversal would leave the e/2 vulnerable to thermal dephasing. The origin of the presence of the e/2 oscillations remains an open question.

In spite of these difficulties, the possibility that e/4 quasiparticles doubly traverse the interferometer offers a model [5, 6] that has importance for its implications as to the statistics of the 5/2 excitations. Here the e/2 oscillations may represent a direct manifestation of the non-Abelian statistics of e/4 quasiparticles. From the theory for non-Abelian e/4 quasiparticles, if the interferometer area A encircled by the e/4 quasiparticle encloses an even number of localized quasiparticles or quasiholes, an e/4 period results: if an odd number of quasiparticles is enclosed, interference and so transmission oscillations are suppressed. Given that enclosure of an odd number of localized quasiparticles suppresses e/4 oscillations due to the non-Abelian statistics, the e/4 quasiparticles can still traverse the perimeter of the interferometer and have the potential to complete two laps around the area A. In this case, the odd number of localized particles in A has now been encircled twice, so that an even number has been encircled in sum, and interference oscillations are now not suppressed. Also, the total area encircled is now twice A, twice the encircled number of magnetic flux quanta, which would produce an effective period of e/2. Because the e/4 period from the single lap is suppressed, the e/2 period from the double lap can be expressed, but at period e/2. This picture is consistent with the sequential observation of e/2 or e/4 periods; as the side gate is swept, an odd or even number of localized quasiparticles may be enclosed. This picture is also consistent with the observation that the e/2 period oscillations are generally of smaller amplitude; due to the longer path length that must be traversed by the quasiparticles over two laps any decoherence mechanisms present will result in a concomitant amplitude reduction. The data of this study do not provide sufficient evidence to conclude this as the proper model. However, this picture is important in considering interferometric determination of potential non-Abelian e/4 excitation statistics. Further detailed experimental studies exploring this possible scenario are presently underway.

In conclusion, the principal findings of this work are that interferometric measurements can be accomplished on the 5/2 FQH state: apparent Aharonov-Bohm oscillations are observed demonstrating periods consistent with e/4 charge for lowest temperatures tested, but also demonstrating oscillation periods consistent with e/2. The origin of the e/2 oscillations is an open question. These findings of e/4 and e/2 periods may be considered in the model of non-Abelian properties of the e/4 quasiparticle.

Acknowledgements:

The authors gratefully acknowledge useful discussions with B.I. Halperin, A. Stern and H.L. Stormer, and acknowledge technical assistance from M. Peabody.

References:

[1] Willett, R. L., Eisenstein, J. P., Stormer, H. L., Tsui, D. C., Gossard, A. C. & English, J. H. Observation of an even-denominator quantum number in the fractional quantum Hall effect. Phys. Rev. Lett. **59**, 1776-1779 (1987).

[2] Moore, G. & Read, N. Nonabelions In The Fractional Quantum Hall Effect. Nucl. Phys. B **360**, 362-396 (1991).

[3] Haldane, F. D. M. & Rezayi, E. H. Spin-singlet wave function for the half-integral quantum Hall effect. Phys. Rev. Lett. **60**, 956-959 (1988).

[4] Kitaev, A. Y. Fault tolerant quantum computation by anyons. Ann. Phys. **303**, 2-30 (2003).

[5] Das Sarma, S., Freedman, M. & Nayak, C. Topologically Protected Qubits from a Possible Non-Abelian Fractional Quantum Hall State. Phys. Rev. Lett. **94**, 166802-166805 (2005).

[6] Stern, A. & Halperin, B. I. Proposed Experiments to Probe the Non-Abelian $\nu$=5/2 Quantum Hall State. Phys. Rev. Lett. **96**, 016802-016805 (2006).

[7] Bonderson, P., Kitaev, A. & Shtengel, K. Detecting non-Abelian statistics in the $\nu$=5/2 fractional quantum Hall state. Phys. Rev. Lett. **96**, 016803-016806 (2006).

[8] C. de C. Chamon, D.S. Freed, S.A Kivelson, S.L. Sondhi, and X-G. Wen, Two point-contact interferometer for quantum Hall systems, Phys. Rev. B **55**, 2331-2343 (1997).

[9] E.Fradkin, Chetan Nayak, A. Tsvelik, and Frank Wilczek, A Chern-Simons Effective Field Theory for the Pfaffian Quantum Hall State, Nucl. Phys. B**516**, 704-718 (1998).

[10] van Loosdrecht, P. et al. Aharonov-Bohm effect in a singly connected point contact. Phys. Rev. B **38**, 10162-10165 (1988).

[11] van Wees, B. J. et al. Observation of Zero-Dimensional States in a One-Dimensional Electron Interferometer. Phys. Rev. Lett. **62**, 2523-2526 (1989).

[12] Ji, Y., Chung, Y., Sprinzak, D., Heiblum, M., Mahalu, D. & Shtrikman, H. An electronic Mach-Zehnder interferometer. Nature **422**, 415-418 (2003).

[13] Roulleau, P., Portier, F., Roche, P., Cavanna, A., Faini, G., Gennser, U. & Mailly, D. Direct Measurement of the Coherence Length of Edge States in the Integer Quantum Hall Regime. Phys. Rev. Lett. **100**, 126802-126805 (2008).

[14] Camino, F. E., Zhou, W. & Goldman, V. J. Aharonov-Bohm Superperiod in a Laughlin Quasiparticle Interferometer. Phys. Rev. Lett. **95**, 246802-246805 (2005), and references therein.

[15] Godfrey, M. D. et al. Aharonov-Bohm-Like Oscillations in Quantum Hall Corrals. Preprint at http://arxiv.org/abs/cond-mat/07082448 (2007).

[16] Willett, R. L., Manfra, M. J., Pfeiffer, L. N. & West, K. W. Confinement of fractional quantum Hall states in narrow conducting channels. Appl. Phys. Lett. **91**, 052105-052107 (2007).

[17] Rosenow, B. & Halperin, B. I. Influence of Interactions on Flux and Back-Gate Period of Quantum Hall Interferometers. Phys. Rev. Lett. **98**, 106801-106804 (2007).

[18] Composite Fermions: A Unified View of the Quantum Hall Regime, edited by O. Heinonen, World Scientific, Singapore, 1998.

[19] Beenaker, C. W. J. & van Houten, H. Quantum Transport in Semiconductor Nanostructures. Solid State Physics **44**, 1-128 (1991).

[20] Dolev, M., Heiblum, M., Umansky, V., Stern, A. & Mahalu, D. Observation of a quarter of an electron charge at the ν=5/2 quantum Hall state. Nature **452**, 829-834 (2008).
[21] Radu, I., Miller, J. B., Marcus, C. M., Kastner, M. A., Pfeiffer, L. N. & West, K. W. Quasi-Particle Properties from Tunneling in the ν=5/2 Fractional Quantum Hall State. Science 320, 899-902 (2008).
[22] see Wan, X., Hu, Z., Rezayi, E., and Yang, K., Fractional quantum Hall effect at ν=5/2: Ground states, non-Abelian quasiholes, and edge modes in a microscopic model, PRB **77**, 165316 (2008).

Figure Legends:

FIGURE 1. Interference device and bulk sample transport: (a) schematic of the interferometer showing ohmic contacts to the 2D electron layer labeled 1 to 4, the confined area $A_L$ delineated by quantum point contacts (qpcs) defined by gates a and c with the central channel controlled by gate b. The side gate b voltage is referred to in these experiments as Vs. Longitudinal resistance ($R_L$) through the device could be measured by the voltage drop from 3 to 4 while driving current from 1 to 2, diagonal resistance ($R_D$) measured from 1 to 4 while driving current from 2 to 3. Example propagating edge states are marked by the arrowed lines along the gates and sample edges, with the dashed lines marking where backscattering may occur at the qpcs. (b) electron micrograph of an interferometer gate structure used in this study. (c) bulk transport at 30mK near but not through an interferometer device on a sample used in this study.

FIGURE 2. Interferometer Aharonov-Bohm (A-B) oscillations: General comparison of the interferometer longitudinal resistance $R_L$ for two types of measurement in which either: the B-field is swept and resistance $R_L$ through the device is measured, or the side gate voltage Vs is swept while $R_L$ is monitored. Inset to (a): overview of $R_L$ ranging from filling factor ν=3 to near ν=3/2 with qpcs at -4.5V bias and central channel bias at -6.0V, temperature=30mK, current 5nA. The box near 90kG roughly delineates the B-field range shown in the top panel. (panel a) Swept B-field A-B oscillations: $R_L$ features consistent with Aharonov-Bohm oscillations through the device with background resistance subtracted to better display the oscillations. Note runs of several oscillations, marked by the vertical lines, each with period ~ 230G. Such runs of oscillations, with similar period, can be observed under these gating and temperature conditions throughout the magnetic field range shown in the inset $R_L$ vs. B-field plot. (panel b) Swept side-gate A-B oscillations: $R_L$ is monitored while sweeping the side gate voltage Vs at fixed B-field. Near ν=2 runs or series of oscillations are present, with disruptions of phase between the runs. The period is marked by vertical lines reflecting the charge, period P~ΔVs~ΔA~(h/e*)/B. Each series or run of coherent oscillations is marked by the charge value. The lower trace shows a similar measurement near ν=5/3, with larger period oscillations. The vertical lines marked on that trace are the period calculated from the ν=2 period assuming e*(ν=2)=e and e*(ν=5/3)=e/3, and scaling the different B-field values: P(ν=5/3)=P(ν=2)(e*(ν=2)/e*(ν=5/3))(B(ν=2)/B(ν=5/3)). The principal oscillations in the ν=5/3 trace agree well with this scaling, consistent with interferometric derivation of the fractional charge e/3 at that filling factor. (panel c) Quantitative verification of the predominant periods in swept area measurements by Fast-Fourier Transforms (FFT). Left and right FFTs correspond to transforms applied to the data in panels b and d respectively. The principal peak frequency for each filling factor is marked by a vertical line with that frequency matched to the vertical period markers in the data of panels b and d. (panel d) From a different sample cool-down, a second example of A-B oscillations near ν=2 (top trace) in which the period of those oscillations is used to determine the oscillation period for A-B oscillations at ν=7/3, matching well those features shown in the lower trace. Temperature is 25mK, current 2nA. Again, each run of coherent oscillations is marked by the charge, three coherent series for ν=2 and three for ν=7/3.

FIGURE 3. A-B interferometric measurements at 5/2 filling factor displaying oscillations with periods corresponding to e/4 and e/2. $R_L$ is measured for side gate sweeps at filling factor ν=5/2 using similar measurements at ν=2 and 5/3 or 7/3 as metrics. Left and right panel columns are data at 25mK, achieved with two different cool-downs from room temperature to 25mK and using current of 2nA. Both sides show interference effects at ν=2, 5/3 or 7/3, and 5/2, with the predominant ν=2 period marked in both by vertical lines. As in Figure 2, this period is used to derive the period expected at 5/3 or 7/3 and additionally at 5/2 using P~ΔVs~(h/e*)/B, assuming the quasiparticle charge of e/3 at 5/3 and 7/3, and e/4 at 5/2. Those expected periods are then marked by vertical lines in the respective traces. In both sample preparations the period of the ν=2 traces determines an e/3 quasiparticle period for the ν=5/3 or 7/3 measurement that is consistent with the predominant oscillation features in each trace. Each run or series of coherent oscillations is marked by the corresponding charge. The bottom panels in each column show fast Fourier transforms (FFTs) of the data in the above panels, showing frequency peaks that verify the marked period lines in the swept gate data.

The bottom traces of each panel are taken near 5/2 filling and both show series of large period oscillations. These prominent features are runs of oscillations with period consistent with quasiparticles of charge e/4 as derived from the period of the ν=2 traces; the marked vertical lines are the e/4 period derived from ν=2 and consistent with peaks of the FFT spectra. In both 5/2 traces additional oscillation features of shorter period are present that correspond to a period as expected from charge e/2, again marked by corresponding vertical lines. The amplitudes of the e/2 oscillations are generally smaller than those of the e/4, thirds and integral filling factors. The bottom panels show FFT spectra of the 5/2 traces, each demonstrating both e/4 and e/2 frequency peaks. Again the arrows show the frequency/period marked in the swept gate data.

FIGURE 4. Quasiparticle interference at 5/2; e/4 and e/2 period oscillations and their properties. Representative data are shown demonstrating the presence and properties of oscillations near filling factor 5/2 in $R_L$ vs. change in side gate voltage Vs measurements consistent with both e/2 and e/4 period oscillations. Top panels (a and b) demonstrate in two different sample preparations that at 29mK within a single side-gate sweep, runs of oscillations with period corresponding to e/2 can be observed sequentially in Vs sweep with oscillations of period consistent with e/4, data similar to the bottom traces of Figure 3. The e/4 and e/2 periods are verified by FFT spectra (panel insets) of the swept side gate data, showing frequency peaks corresponding to e/4 and e/2 periods. The marked FFT peaks match the marked vertical period lines in the swept side gate data. Panel (c) data indicate temperature dependence of e/4 and e/2 oscillations: e/2 oscillations may be made more prevalent with an increase in temperature. The temperature of the sample was taken from 30mK to near 150mK during a single side gate voltage sweep with e/4 period oscillations dominant at the low temperature and e/2 oscillations present at the higher temperature. Temperature dependence is further examined in panel d near filling factor

5/2 data from a different overall cool-down of the sample; $R_L$ vs. Vs traces for the same range of Vs are measured at 25mK, 130mK, and 610mK. Data are offset for clarity. The data display increased e/2 oscillations prevalence for the higher temperature (130mK), and loss of oscillatory structure at temperatures higher than those able to support the 5/2 state, 610mK. A current of 2nA is used in panels a, b, and d, 5nA in panel c.

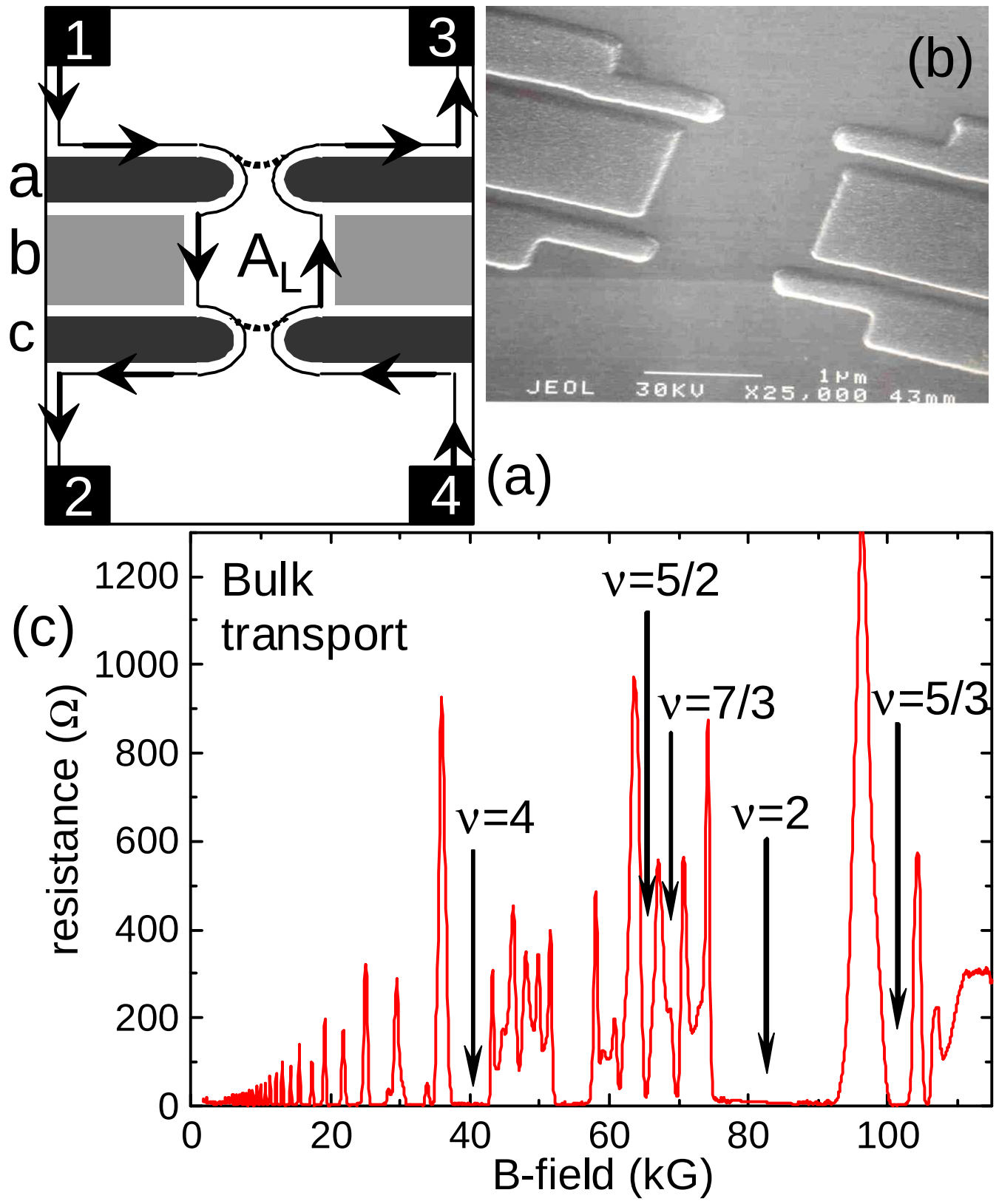

Figure 1.

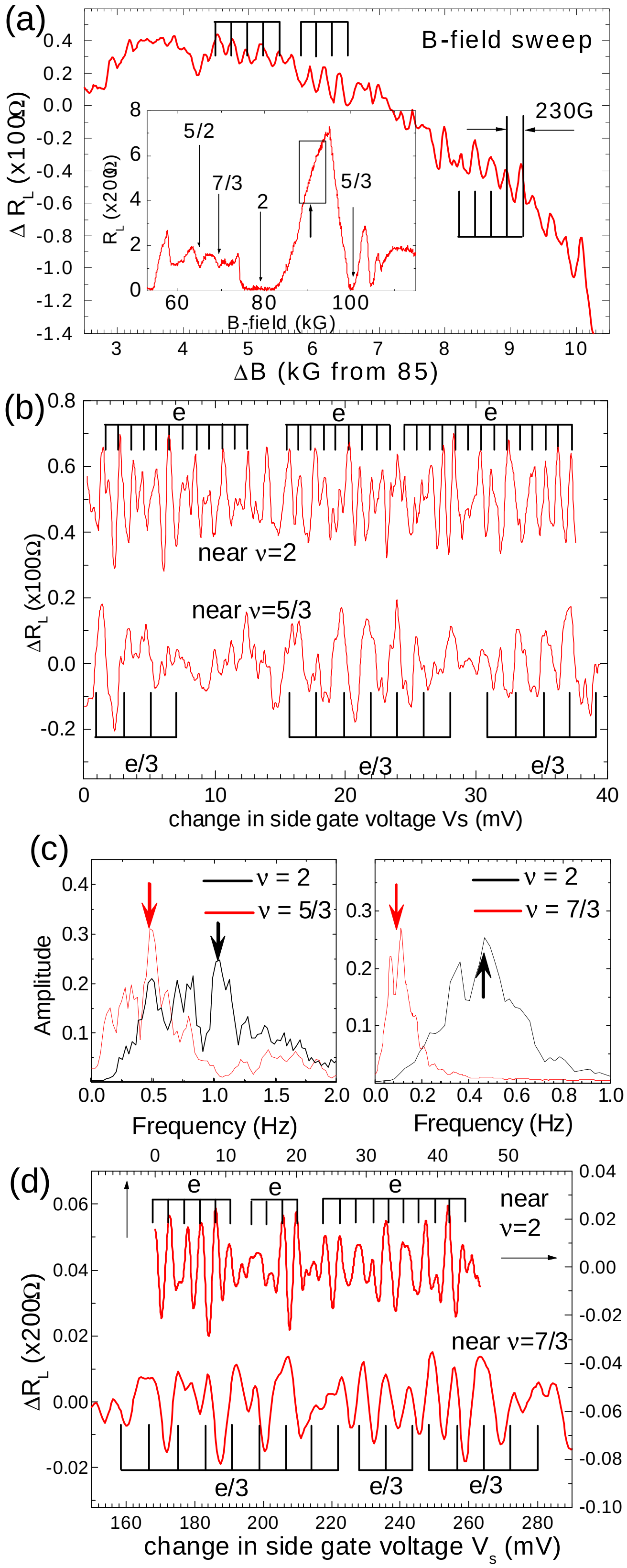

Figure 2.

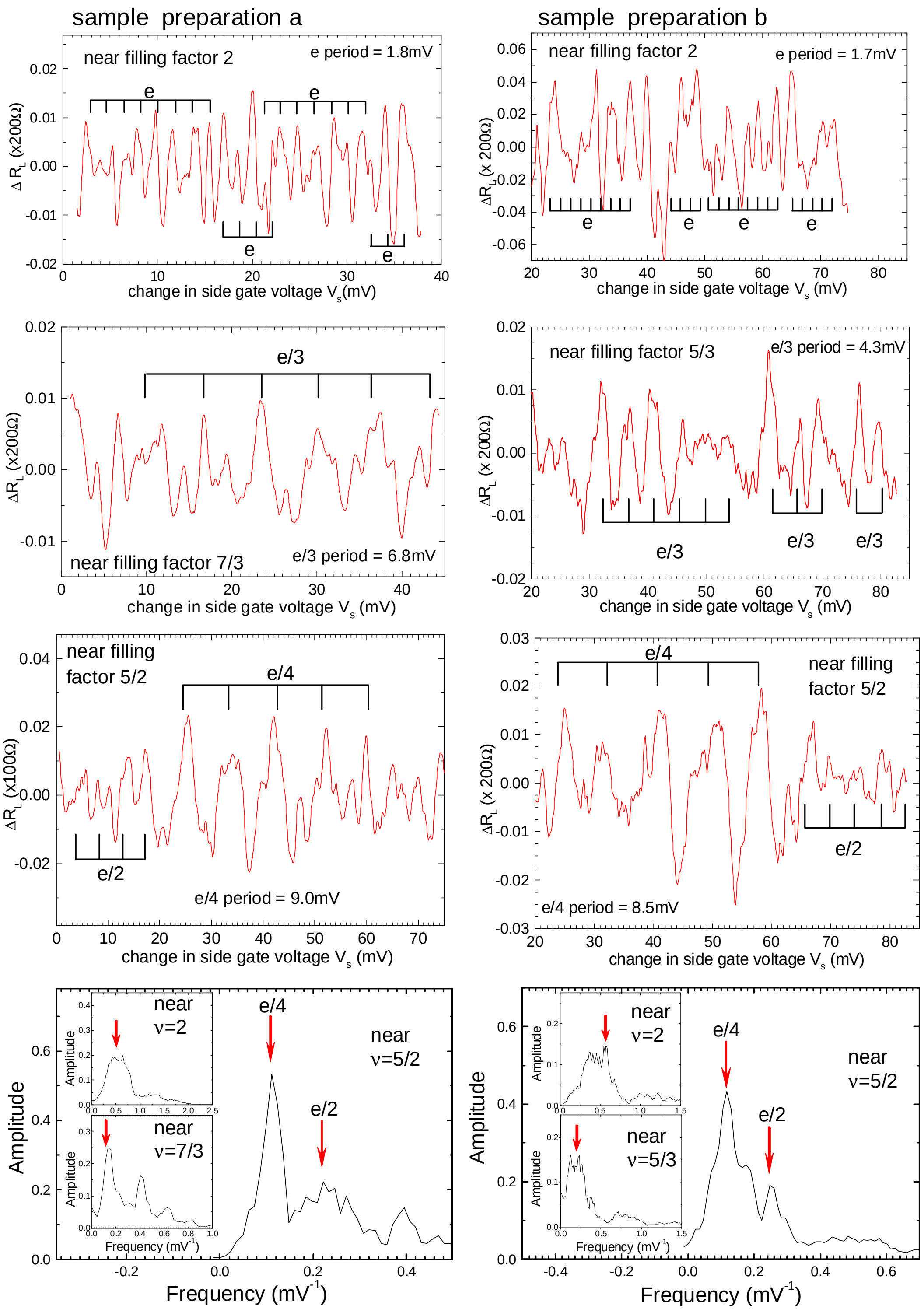

sample preparation a
sample preparation b
near filling factor 2
e period = 1.8mV
e period = 1.7mV
near filling factor 7/3
e/3 period = 6.8mV
near filling factor 5/3
e/3 period = 4.3mV
near filling factor 5/2
e/4 period = 9.0mV
e/4 period = 8.5mV
change in side gate voltage $V_s$ (mV)
Amplitude
Frequency (mV$^{-1}$)
near ν=2
near ν=7/3
near ν=5/3
near ν=5/2
e/4
e/2

Figure 3.

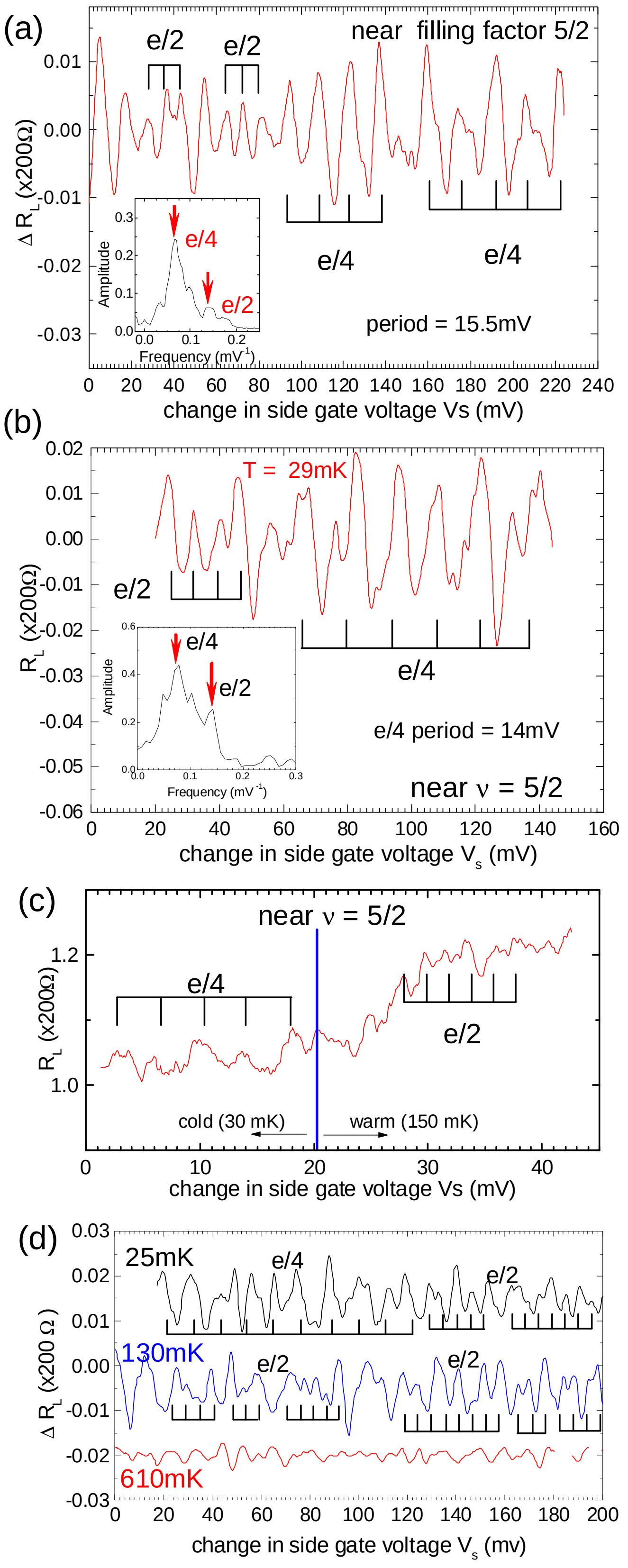
(a)
near filling factor 5/2
e/2
e/2
e/4
e/4
period = 15.5mV
Δ R_L (x200Ω)
Amplitude
Frequency (mV^-1)
change in side gate voltage Vs (mV)
(b)
T = 29mK
e/2
e/4
e/4
e/2
e/4 period = 14mV
near ν = 5/2
R_L (x200Ω)
change in side gate voltage V_s (mV)
(c)
near ν = 5/2
e/4
e/2
cold (30 mK)
warm (150 mK)
R_L (x200Ω)
change in side gate voltage Vs (mV)
(d)
25mK
e/4
e/2
130mK
e/2
e/2
610mK
Δ R_L (x200 Ω)
change in side gate voltage V_s (mv)

Figure 4.

Supporting Information:

Section I:  Methods and device operation:

Experiments were performed on high mobility (>25x10$^{6}$cm$^{2}$/V-sec), high density (~4x10$^{11}$/cm$^{2}$) AlGaAs/GaAs heterstructures. These samples were designed to optimize the 5/2 filling factor state in top gate devices. The 2D electron systems reside roughly 200nm below the sample surface in all samples tested. Sample illumination with a red LED is used to maximize mobility. At the sample surface a large scale (0.6mmx0.3mm) mesa is etched in the sample to provide a Hall bar, with contacting on the perimeter and gates placed internally and extending to the edge of the Hall bar. Before gate definition a thin (<40nm) amorphous SiN layer is deposited on the entire surface. The high mobility of the device is preserved as demonstrated by the transport in main text Figure 1c. The confinement device surface gates, comprised of ~60nm Al, are arranged in the form of two quantum point contacts enclosing a channel (Figure 1a). Each pair of quantum point contacts can be biased independently, as is true for the channel referred to as voltage Vs, allowing independent transmission and size control of the confinement area delineated by the surface gates. The total area as defined by the surface gates is greater than 2μm$^{2}$ (see electron micrograph, Figure 1c). Some of the channel properties and gating results have been described elsewhere [16]. Configurations for resistance measurements [19] corresponding to longitudinal ($R_L$) and diagonal ($R_D$) resistances are described in the text according to Figure 1a schematic. Standard lock-in techniques are employed for these measurements. Low temperatures are achieved using a dilution refrigerator. Gate performance is monitored using both the resistance measurements and gate leakage measurements. $R_L$ values shown in this study are taken for non-discernable gate leakage (<4pA) which is viable to over -6V bias. Extensive gate testing is employed, with these samples demonstrating full depletion at bias of near -2.5V. The three components of the confinement device are first tested for integrity; each qpc and the channel are independently tuned with $R_L$ demonstrating zero B-field diffuse boundary scattering as the signature that the gate is viable [16]. The qpcs, each roughly 1 μm in width, provide tunability crucial to the experiments; testing of the magneto-conductance spectra is carried out to determine the desired transmission properties, focusing on preservation of FQHE features.

Section II:  Comparisons of A-B data from side gate sweeps at $\nu$ = 2, 5/3, and 5/2

Supporting information Figure S1 contains two panels, each demonstrating $R_L$ for side gate sweeps comparing filling factors 2, 5/3, and 5/2. The predominant oscillation period at $\nu$=2 is used to calculate the A-B period that should be observed at 5/3 and 5/2: the marked vertical lines in the $\nu$=2 trace are measured, the vertical lines in the 5/3 and 5/2 traces are the calculated lines. Good correspondence is seen in the 5/3 and 5/2 traces to the periods derived from $\nu$=2. The data cover relatively small side gate sweeps, with excitation current of 5nA in these traces, and refrigerator temperature in these traces less than 30mK. In addition to e/4 oscillations, the data display evidence for e/2 period oscillations.

These data represent that taken in the early cool-downs of these experiments. The sample preparation here included cooling to base temperatures and illumination at that temperature, followed by charging of the gates. In these procedures measurements immediately followed charging of the gates. In the interim it has been observed that allowing the gate charged device to equilibrate over a several day period exposes a lower noise signal that promotes observation of the interference features, such as the e/2 oscillations. It has also been noted that excitation current of 2nA rather than 5nA, as used in the early experimental trials, promotes observation of the interference effects.

The nature of the findings for the side gate sweeps at the various filling factors, that of short runs of oscillations at the respective periods, raises the question of how reproducible are the findings, or what is the success rate at observing the oscillations? Using a scan of the side gate of roughly 40mV or more, in each of the four cool downs sets of the target filling factors (2, 5/3, 7/3 and 5/2) were tested. Oscillations of the appropriate period were observed in *each* of the scans performed for *all* sets tested specifically for optimized qpc voltages and for magnetic field position within a range around the target filling factor. It was found that for qpcs voltages set at only marginally more than depletion or at high values nearing pinch-off, the oscillations were not consistent or were absent. At the optimized qpc voltages and magnetic field values repeated side gate scans showed runs of oscillations in each scan that are not specifically reproducible *in detail* in the number of oscillations in each run, but with sufficiently long gate equilibration the general features of the oscillatory sets or runs can be reproduced.

When examining 5/2 filling factor, as shown above, oscillations with periods consistent with e/2 and/or e/4 charge were observed in each of the scans at lowest temperatures, again under the optimized conditions. While the data shown above and collected to date demonstrate a high prevalence of e/4 consistent oscillations, often coexistent in side gate sweeps with e/2 oscillations, a precise determination of that relative prevalence is presently under study.

## Section III. A-B oscillations at 5/2 with periods e/4 and e/2

Supporting information Figures S2a and S2b demonstrate data sets of $R_L$ versus side gate voltage near filling factor 5/2 and at base temperature near 25mK for two different sample preparations using 2nA current. A-B oscillations features representing e/4 and e/2 periods are apparent in both traces. The vertical lines delineating those periods are derived from oscillations measured at filling factor 2 for each and are consistent with FFT spectra of the traces. As noted in the text, the e/2 oscillations are generally of smaller amplitude than the e/4 oscillations.

The marking of e/4 and e/2 periods are supported by fast Fourier transforms of the spectra, each showing peak frequencies consistent with the marked periods in the side gate voltage sweep data. Separation of the data into sections of e/2 and e/4 periods is shown in the lower panels, where FFT spectra of those separated data are also displayed. The peaks in the FFT spectra of the separated data further demonstrate the independent occurrence of the two periods e/4 and e/2.

## Section IV. Temperature dependence at 5/2 with 5nA drive current

Supporting information Figure S3 shows $R_L$ versus side gate sweep data near 5/2 filling factor at three different temperatures. At the lowest temperature of 25mK structures are present at period consistent with charge e/4, with the period marked and Vs scaling accomplished as described previously with respect to filling factor 2. At the higher temperature of 76mK periodic features corresponding to e/2 are apparent and fill the Vs scan window. At 145mK little $R_L$ structure is present, with only a hint of e/2 properties.

This data set resulted from measurements undertaken early in the execution of the experiments using a relatively high drive current of 5nA. As noise reduction within the experiment was accomplished, and sample preparation became more refined as described above, lower drive current of 2nA could be used. As employed in the data of Figure 4d and many of the prior Figures, this lower current may have reduced heating within the interference device compared to using 5nA . Given this point, the data in Figures S1 and S3 may represent a functionally higher temperature in the device than the temperature posted, and possibly higher than that using 2nA current. The higher resolution of the interference features in the more recent experimental results, those using the lower current of 2nA, can be attributed to this noise reduction and may also reflect the expression of the interference at a lower temperature within the device.

Supporting Information Figure Captions:

Figure S1. Interferometer longitudinal resistance versus side gate voltage change measured at multiple filling factors for two sample cool-downs. Excitation current is 5nA, temperature is 25mK.

Figures S2a and b. Interferometer longitudinal resistance versus side gate voltage change operated near filling factor 5/2 for two different sample illuminations. Excitation current is 2nA, temperature is 25mK. Fast Fourier Transforms (FFTs) of the entire side gate sweep spectra are shown within the top panels as insets, both demonstrating peaks at frequencies corresponding to e/4 and e/2 periods. The lower panels show the sections of e/4 and e/2 oscillations that are each taken from the top panel. FFTs are then performed on those sections, showing their respective peak frequencies. In both cases of the isolated e/2 oscillations the marked period in $R_L$ versus gate voltage change matches well the peak of the FFT spectrum.

Figure S3. Interferometer longitudinal resistance versus side gate voltage change measured near filling factor 5/2 at three different temperatures. Excitation current is

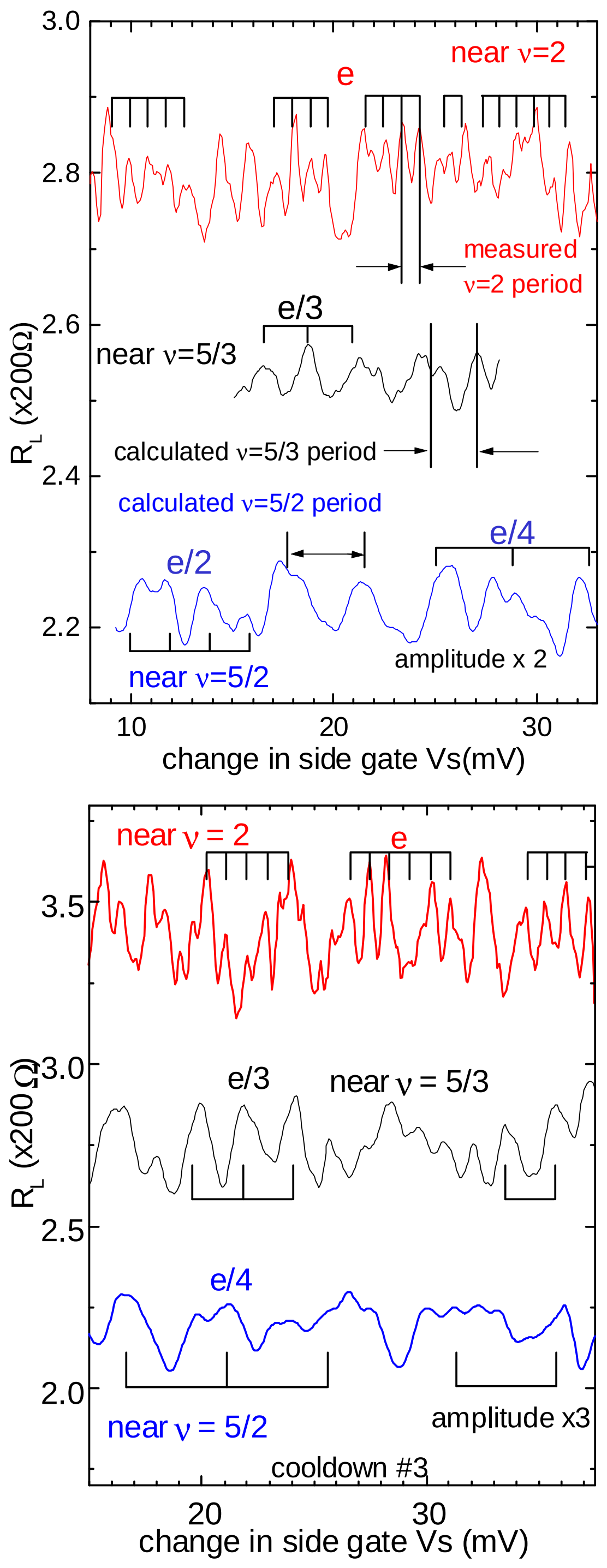

Figure S1.

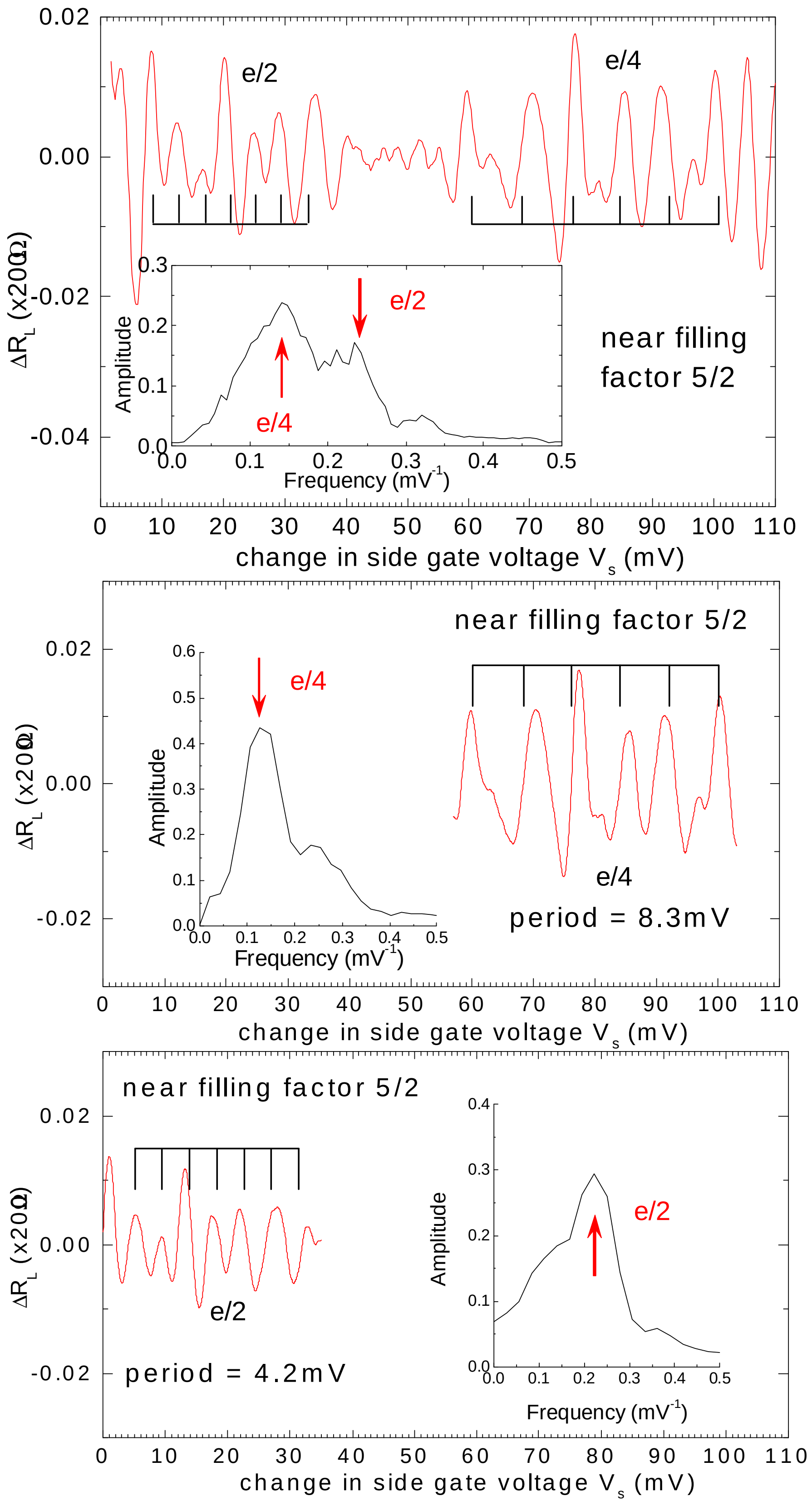

Figure S2a.

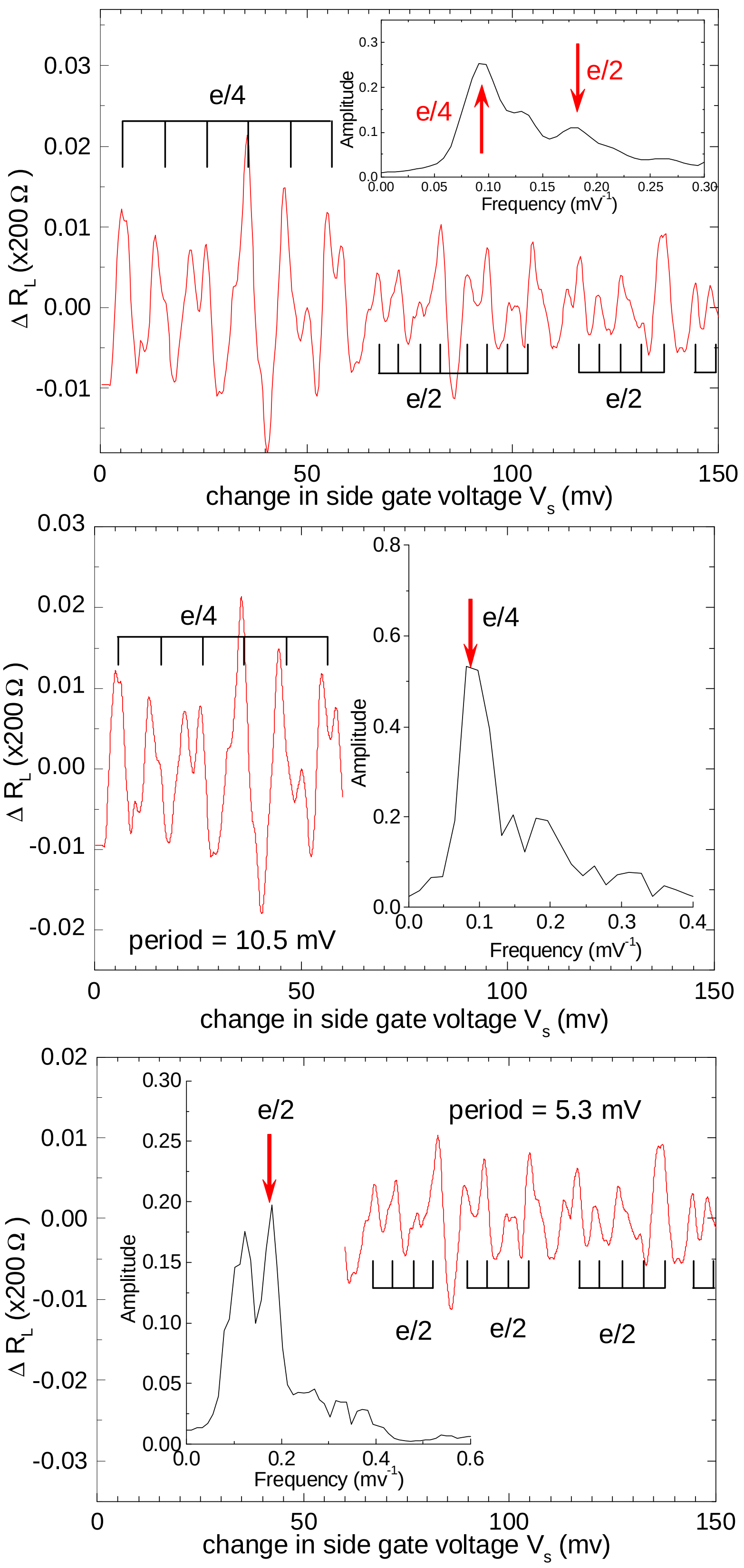
e/4
e/2
Amplitude
Frequency (mV^-1)
e/2
e/2
Δ R_L (x200 Ω )
change in side gate voltage V_s (mv)
period = 10.5 mV
period = 5.3 mV
Frequency (mv^-1)

Figure S2b.

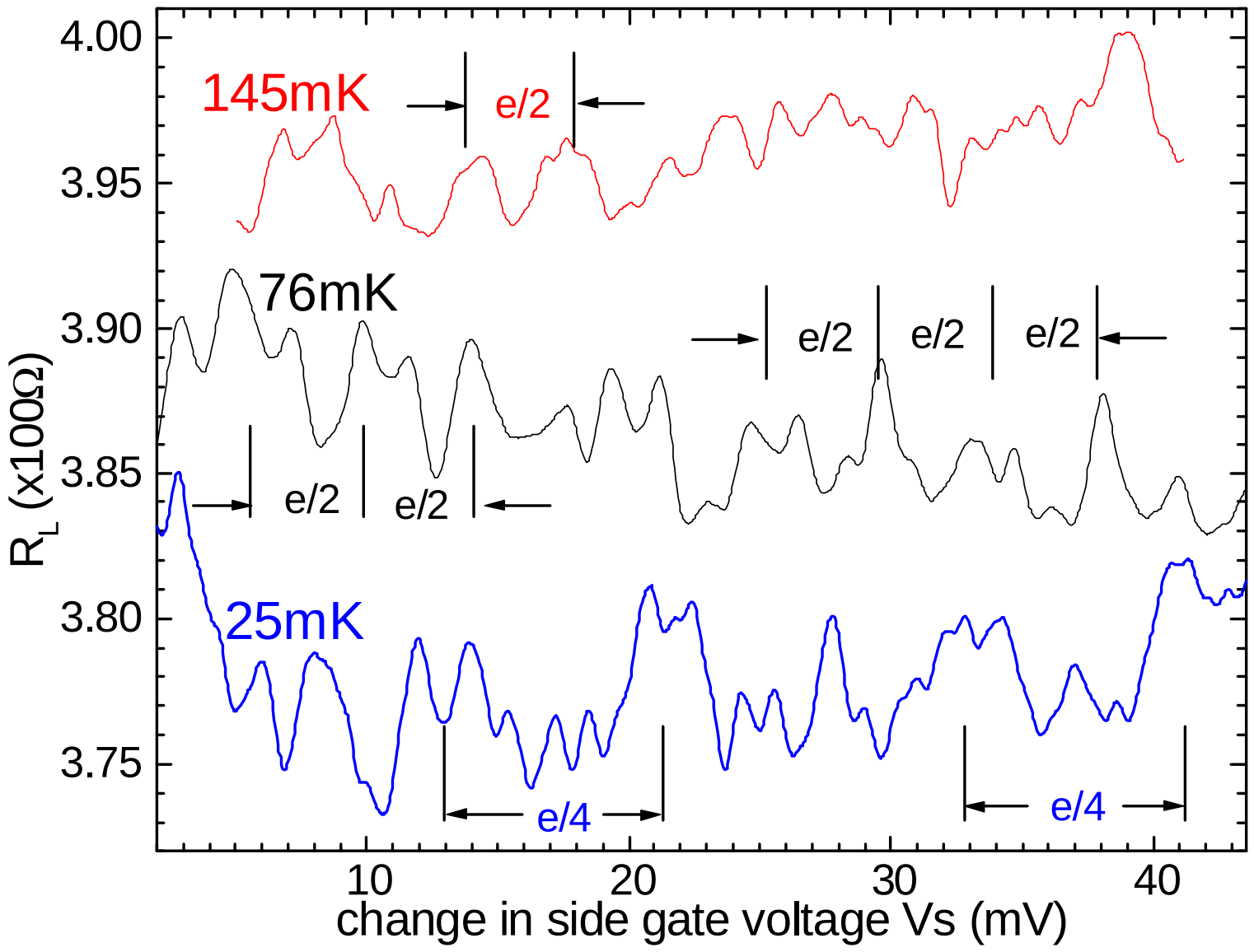

Figure S3.